\documentclass[aps,prc,preprint,amsmath,showpacs,superscriptaddress,nofootinbib]{revtex4}
\usepackage{graphicx,epsfig}
\newcommand{\beqy}{\begin{eqnarray}}
\newcommand{\eeqy}{\end{eqnarray}}
\newcommand{\bmlet}{\begin{subequations}}
\newcommand{\emlet}{\end{subequations}}

\begin{document}

\title{Pairing: from atomic nuclei to neutron-star crusts}

\author{N.~Chamel}
\affiliation{Institut d'Astronomie et d'Astrophysique, CP-226, Universit\'e
Libre de Bruxelles, 1050 Brussels, Belgium}
\author{J.~M.~Pearson}
\affiliation{D\'ept. de Physique, Universit\'e de Montr\'eal, Montr\'eal 
(Qu\'ebec), H3C 3J7 Canada}
\author{S.~Goriely}
\affiliation{Institut d'Astronomie et d'Astrophysique, CP-226,
Universit\'e Libre de Bruxelles, 1050 Brussels, Belgium}

\begin{abstract}
Nuclear pairing is studied both in atomic nuclei and in neutron-star crusts
in the unified framework of the energy-density functional theory using generalized 
Skyrme functionals complemented with a local pairing functional obtained from 
many-body calculations in homogeneous nuclear matter using realistic forces. 
\end{abstract}

\maketitle

\section{Introduction}

The possibility of pairing in atomic nuclei was first studied by Bohr, 
Mottelson and Pines~\cite{bohr58} and by Belyaev~\cite{bel59} only one year after the publication of the theory 
of superconductivity by Bardeen, Cooper and Schrieffer (BCS)~\cite{bcs57}. Meanwhile, Bogoliubov developed a 
microscopic theory of superfluidity and superconductivity and explored its consequences
for nuclear matter~\cite{bog58}. In 1959, Migdal speculated that the interior of neutron stars 
might be superfluid~\cite{mig59} and this scenario was further investigated by Ginzburg 
and Kirzhnits in 1964~\cite{gk64}. Soon after the discovery of the first pulsars, the observation of 
frequency glitches followed by very long relaxation times of the order of months provided 
strong evidence of nuclear superfluidity~\cite{baym69}. Pulsar glitches are believed to be 
related to the dynamics of the neutron superfluid permeating the inner layers of the solid 
neutron star crust~\cite{and75}. Superfluidity plays also a predominant role in
neutron-star cooling (see Page in this volume).

The pairing phenomenon in both finite systems like atomic nuclei and in 
infinite nuclear matter can be consistently described in the framework of the 
energy-density functional (EDF) theory (see Dobaczewski and Nazarewicz in this 
volume). This theory, which has been historically formulated in terms of 
effective interactions in the context of self-consistent mean-field methods, 
has been 
very successful in describing the structure and the dynamics of a wide range of nuclei~\cite{bhr03}. 
These interactions have been also commonly applied to the modeling of neutron-star interiors. Actually 
no sooner did Skyrme~\cite{sky59} introduce his eponymous effective interaction than Cameron~\cite{cam59} 
applied it to calculate the structure of neutron stars. By showing that their maximum mass was significantly 
higher than the Chandrasekhar mass limit, his work brought support to the scenario of neutron star formation 
from the catastrophic gravitational core-collapse of massive stars during type II supernova explosions, as 
proposed much earlier by Baade and Zwicky~\cite{bz33}.

\section{Nuclear energy density functional theory in a nutshell}

Assuming time-reversal symmetry, the ground-state energy $E$ is supposed to depend on 
(i) the nucleon density (denoting the spin states by $\sigma=\pm1$ and $q = n$ 
or $p$ for neutron or proton, respectively),
\beqy
\label{2}
\rho_q(\pmb{r}) = \sum_{\sigma=\pm 1}\rho_q(\pmb{r}, \sigma; \pmb{r}, \sigma)
\, ,
\eeqy
(ii) the kinetic-energy density (in units of $\hbar^2/2M_q$ where $M_q$ is the nucleon mass),
\beqy
\label{3}
\tau_q(\pmb{r}) = \sum_{\sigma=\pm 1}\int\,{\rm d}^3\pmb{r^\prime}\,\delta(\pmb{r}-\pmb{r^\prime}) \pmb{\nabla}\cdot\pmb{\nabla^\prime}
\rho_q(\pmb{r}, \sigma; \pmb{r^\prime}, \sigma)  \, ,
\eeqy
(iii) the spin-current density,
\beqy
\label{4}
\pmb{J}_q(\pmb{r}) = -{\rm i}\sum_{\sigma,\sigma^\prime=\pm1}\int\,{\rm d}^3\pmb{r^\prime}\,\delta(\pmb{r}-\pmb{r^\prime})
\pmb{\nabla}\rho_q(\pmb{r}, \sigma; \pmb{r^\prime},
\sigma^\prime) \times \pmb{\hat\sigma}_{\sigma^\prime \sigma}
\eeqy
and (iv) the abnormal density,
\beqy
\label{5}
\tilde{\rho}_q(\pmb{r}) = \sum_{\sigma=\pm 1}
\tilde{\rho}_q(\pmb{r}, \sigma ; \pmb{r}, \sigma)   \, ,
\eeqy
where $\pmb{\hat\sigma}_{\sigma\sigma^\prime}$ denotes the Pauli spin
matrices. In turn the normal and abnormal density matrices,
$\rho(\pmb{r}, \sigma; \pmb{r^\prime}, \sigma^\prime)$ and
$\tilde{\rho}(\pmb{r}, \sigma; \pmb{r^\prime}, \sigma^\prime)$ respectively,
can be expressed as
\beqy
\label{6}
\rho_q(\pmb{r}, \sigma; \pmb{r^\prime}, \sigma^\prime) =
\sum_{i(q)}\psi^{(q)}_{2i}(\pmb{r}, \sigma)\psi^{(q)}_{2i}(\pmb{r^\prime}, \sigma^\prime)^*
\eeqy
and
\beqy
\label{7}
\tilde{\rho}_q(\pmb{r}, \sigma; \pmb{r^\prime}, \sigma^\prime) =
-\sum_{i(q)}\psi^{(q)}_{1i}(\pmb{r}, \sigma)\psi^{(q)}_{2i}(\pmb{r^\prime},
\sigma^\prime)^* \, ,
\eeqy
where $\psi^{(q)}_{1i}(\pmb{r}, \sigma)$ and $\psi^{(q)}_{2i}(\pmb{r}, \sigma)$
are the two components of the quasiparticle (q.p.) wavefunction.
Here, as throughout 
this paper pure nucleon states are being assumed; the more general formalism 
involving neutron-proton mixing has been developed in Ref.~\cite{perl04}. 

Minimizing the total energy $E$ with respect to $\psi^{(q)}_{1i}(\pmb{r}, \sigma)$ and 
$\psi^{(q)}_{2i}(\pmb{r}, \sigma)$ under the constraints of fixed particle numbers 
leads to the Hartree-Fock-Bogoliubov (HFB) equations\footnote{These equations are 
also called Bogoliubov-de Gennes equations in condensed matter physics.}
(see Dobaczewski and Nazarewicz in this volume)
\beqy
\label{8}
\sum_{\sigma^\prime}
\begin{pmatrix} h_q(\pmb{r} )_{\sigma \sigma^\prime} & \Delta_q(\pmb{r}) \delta_{\sigma \sigma^\prime} \\ \Delta_q(\pmb{r}) \delta_{\sigma \sigma^\prime} & -h_q(\pmb{r})_{\sigma \sigma^\prime} 
 \end{pmatrix}\begin{pmatrix} 
\psi^{(q)}_{1i}(\pmb{r},\sigma^\prime) \\ \psi^{(q)}_{2i}(\pmb{r},\sigma^\prime) \end{pmatrix} = \nonumber\\
\begin{pmatrix} E_i+\lambda_q & 0 \\ 0 & E_i-\lambda_q \end{pmatrix}
\begin{pmatrix} \psi^{(q)}_{1i}(\pmb{r},\sigma) \\ \psi^{(q)}_{2i}(\pmb{r},\sigma) \end{pmatrix}
\eeqy
where $\lambda_q$ are Lagrange multipliers. 
The single-particle (s.p.) Hamiltonian 
$h_q(\pmb{r} )_{\sigma \sigma^\prime}$ is given by 
\beqy
\label{9}
h_q(\pmb{r})_{\sigma^\prime\sigma} \equiv -\pmb{\nabla}\cdot
B_q(\pmb{r})\pmb{\nabla}\, \delta_{\sigma\sigma^\prime}
+ U_q(\pmb{r}) \delta_{\sigma\sigma^\prime}
-{\rm i}\pmb{W_q}(\pmb{r}) \cdot\pmb{\nabla}\times\pmb{\hat\sigma}_{\sigma^\prime\sigma}
\eeqy
with the s.p. fields defined by the functional derivatives
\beqy
\label{10}
B_q(\pmb{r}) &=&
\frac{\delta E}{\delta\tau_q(\pmb{r})}\, ,
\hskip0.5cm
U_q(\pmb{r})=\frac{\delta E}{\delta \rho_q(\pmb{r})}\, ,
\hskip0.5cm
\pmb{W}_q(\pmb{r})=\frac{\delta E}{\delta \pmb{J}_q(\pmb{r})}  \, .
\eeqy
Using a local pairing EDF of the form
\beqy
\label{11}
\mathcal{E}_{\rm pair}(\pmb{r})=\frac{1}{4} \sum_{q=n,p} v^{\pi\, q} [\rho_n(\pmb{r}),\rho_p(\pmb{r})] \tilde{\rho}_q(\pmb{r})^2 \, .
\eeqy
the pairing field is given by 
\beqy
\label{12}
\Delta_q(\pmb{r})=\frac{\delta E}{\delta \tilde{\rho}_q(\pmb{r})}=\frac{1}{2}v^{\pi q} [\rho_n(\pmb{r}),\rho_p(\pmb{r})]\tilde{\rho}_q(\pmb{r}) \, .
\eeqy
Expressions for these fields can be found for instance in 
Refs.~\cite{cha08,cgp09}. 
Eqs.~(\ref{8})-(\ref{12}) are still valid at finite temperatures, but 
Eqs. ~(\ref{6}) and (\ref{7}) will have to be replaced by
\beqy
\label{6b}
\rho_q(\pmb{r}, \sigma; \pmb{r^\prime}, \sigma^\prime) =
\sum_{i(q)}f_i\psi^{(q)}_{1i}(\pmb{r}, \sigma)\psi^{(q)}_{1i}(\pmb{r^\prime}, \sigma^\prime)^* \nonumber \\
+ (1-f_i)\psi^{(q)}_{2i}(\pmb{r}, \sigma)\psi^{(q)}_{2i}(\pmb{r^\prime}, \sigma^\prime)^* 
\eeqy
and
\beqy
\label{7b}
\tilde{\rho}_q(\pmb{r}, \sigma; \pmb{r^\prime}, \sigma^\prime) =
\sum_{i(q)}(2f_i-1)\psi^{(q)}_{2i}(\pmb{r}, \sigma)
\psi^{(q)}_{1i}(\pmb{r^\prime}, \sigma^\prime)^* \, ,
\eeqy
where $f_i$ are the q.p. occupation probabilities given by (setting the Boltzmann 
constant  $k_{\rm B}=1$) 
\beqy
f_i=\frac{1}{1+\exp(E_i/T)}\quad .
\eeqy

\section{Skyrme functionals}

The nuclear EDF that we consider here is of the Skyrme type~\cite{bhr03}, i.e.,
\beqy
\label{1}
E=E_{\rm kin}+E_{\rm Coul}+E_{\rm Sky}+E_{\rm pair}\quad ,
\eeqy
where $E_{\rm kin}$ is the kinetic energy of the normalization volume, $E_{\rm Coul}$ is the Coulomb energy
(dropping the exchange part in order to simulate neglected effects such as
Coulomb correlations, charge-symmetry breaking of the nuclear forces and vacuum polarization as discussed in Ref.~\cite{gp08}), 
$E_{\rm Sky}$ is the Skyrme energy and $E_{\rm pair}$ is the nuclear 
pairing energy.

Historically the Skyrme energy was obtained from the Hartree-Fock 
approximation using an effective interaction of the form~\cite{bhr03}
\beqy
\label{13}
v^{\rm Sky}_{i,j} & = & 
t_0(1+x_0 P_\sigma)\delta({\pmb{r}_{ij}})
+\frac{1}{2} t_1(1+x_1 P_\sigma)\frac{1}{\hbar^2}\left[p_{ij}^2\,
\delta({\pmb{r}_{ij}}) +\delta({\pmb{r}_{ij}})\, p_{ij}^2 \right]\nonumber\\
& &+t_2(1+x_2 P_\sigma)\frac{1}{\hbar^2}\pmb{p}_{ij}.\delta(\pmb{r}_{ij})\,
 \pmb{p}_{ij}
+\frac{1}{6}t_3(1+x_3 P_\sigma)\rho(\pmb{r})^\alpha\,\delta(\pmb{r}_{ij})
\nonumber\\
& &+\frac{\rm i}{\hbar^2}W_0(\pmb{\hat\sigma}_i+\pmb{\hat\sigma}_j)\cdot
\pmb{p}_{ij}\times\delta(\pmb{r}_{ij})\,\pmb{p}_{ij}  \quad ,
\eeqy
where $\pmb{r}_{ij} = \pmb{r}_i - \pmb{r}_j$, $\pmb{r} = (\pmb{r}_i + 
\pmb{r}_j)/2$, $\pmb{p}_{ij} = - {\rm i}\hbar(\pmb{\nabla}_i-\pmb{\nabla}_j)/2$
is the relative momentum, $P_\sigma$ is the two-body spin-exchange 
operator. Likewise, the pairing energy can be obtained from a Skyrme-like effective 
interaction given by 
\beqy
\label{14}
v^{\rm pair}_{i, j}= 
\frac{1}{2}(1-P_\sigma)v^{\pi\, q}[\rho_n(\pmb{r}),\rho_p(\pmb{r})]~\delta(\pmb{r}_{ij})\, .
\eeqy
Because of the zero-range of the pairing force, a cutoff has to be used in the gap 
equations in order to avoid divergences (for a review of the various prescriptions, 
see for instance Ref.~\cite{dug05}). 

The density dependence of the pairing strength $v^{\pi\,q}[\rho_n,\rho_p]$
remains very poorly known. It has been usually assumed that it depends only on the isoscalar 
density $\rho=\rho_n+\rho_p$ and has often been parametrized as~\cite{ber91}
\beqy
\label{15}
v^{\pi\,q}[\rho_n,\rho_p] = V_{\pi q}^\Lambda\left(1-\eta_q
\left(\frac{\rho}{\rho_0}\right)^{\alpha_q}\right) \, ,
\eeqy
where $\rho_0$ is the nuclear saturation density while $V_{\pi q}^\Lambda$, $\eta_q$ and $\alpha_q$ are 
adjustable parameters. The superscript $\Lambda$ on $V_{\pi q}^\Lambda$ serves
as a reminder that the 
pairing strength depends very strongly on the cutoff. In principle changing the cutoff 
modifies also the other parameters but the effects are generally found to be small. Effective 
interactions with $\eta_q=0$ ($\eta_q=1$) have been traditionally refered as volume (surface) pairing. 
The standard prescription to fix the parameters is to adjust the value of the pairing strength $V_{\pi q}^\Lambda$ 
to the average gap in $^{120}$Sn~\cite{doba95}. However this does not allow an unambiguous determination of 
the remaining parameters $\eta_q$ and $\alpha_q$. Systematic studies of nuclei seem to favor a so-called mixed 
pairing with $\eta_q\sim 0.5$ and $1/2\lesssim\alpha_q\lesssim 1$~\cite{doba01,sam03}. 

The parameters of the Skyrme EDF are usually determined so as to reproduce a 
set of nuclear data selected according to a specific purpose. The 
non-uniqueness of the fitting procedure has led to a large number of different
parametrizations. Some of them may yield very different 
predictions when applied outside the domain where they were fitted. 
This situation is particularly unsatisfactory for nuclear astrophysical applications which 
require the knowledge of nuclear masses for nuclei so neutron rich that there is no hope of 
measuring them in the foreseeable future; such nuclei play a vital role in the r-process of 
nucleosynthesis~\cite{arn07} and are also found in the outer crust of neutron stars~\cite{pear09}. 
Extrapolations far beyond the neutron drip line are required for the 
description of the inner crust of neutron stars~\cite{onsi08} where nuclear clusters are 
embedded in a sea of unbound neutrons, which are expected to become superfluid at low 
enough temperatures. 
Even though the crust represents only $\sim 1\%$ of the neutron-star mass, it is intimately related to 
many observed astrophysical phenomena~\cite{lrr}.
The need for more reliable extrapolations of these nuclear EDFs has motivated 
recent efforts to construct non-empirical effective interactions and more 
generally microscopic nuclear EDFs~\cite{drut10}. 
Unfortunately such \textit{ab initio} nuclear EDFs have not yet been developed
to the point where they can reproduce nuclear data with the same degree of 
accuracy as do phenomenological EDFs, which can now fit essentially all 
the nuclear mass data with rms deviations lower than 0.6 MeV~\cite{gcp10}.

\section{Pairing in infinite homogeneous nuclear matter}

In infinite homogeneous matter the HFB equations~(\ref{8}) can be readily solved. The q.p. energies 
are given by 
\beqy
\label{16}
E^{(q)}_k = \sqrt{(\varepsilon^{(q)}_k-\lambda_q)^2+\Delta_q^2} \, ,
\eeqy
in terms of the s.p. energies 
\beqy
\label{17}
\varepsilon^{(q)}_k =B_q k^2 + U_q \, .
\eeqy
The q.p. wavefunctions reduce to 
\beqy
\label{18}
\psi^{(q)}_{1k}(\pmb{r},\sigma) = U^{(q)}_{k}\, \phi_k(\pmb{r},\sigma)  \, , \hskip0.5cm \psi^{(q)}_{2k}(\pmb{r},\sigma)=V_{k}^{(q)}\, \phi_k(\pmb{r},\sigma) \, ,
\eeqy
with 
\beqy
\label{19}
U^{(q)}_{k}= \frac{1}{\sqrt{2}}\left(1+\frac{\varepsilon^{(q)}_k-\lambda_q}{E^{(q)}_k}\right)^{1/2} \, ,
\eeqy
\beqy
\label{20}
V^{(q)}_{k}= \frac{1}{\sqrt{2}}\left(1-\frac{\varepsilon^{(q)}_k-\lambda_q}{E^{(q)}_k}\right)^{1/2} \, ,
\eeqy
and $\phi_k(\pmb{r},\sigma)$ is given by 
\beqy
\label{21}
\phi_k(\pmb{r},\sigma) \equiv \frac{1}{\sqrt{V}}\, \exp\left({\rm i} \pmb{k}\cdot\pmb{r}\right) \chi(\sigma) \, ,
\eeqy
where $\chi(\sigma)$ is the Pauli spinor and $V$ is the normalization volume.
The uniform pairing field obeys the well-known isotropic BCS gap equations (see for instance Appendix B of Ref.~\cite{cha08})
\beqy
\label{22}
\Delta_q=-\frac{1}{2} v^{\pi\, q}[\rho_n,\rho_p] \, \Delta_q \,\int_\Lambda {\rm d}\varepsilon 
\frac{g(\varepsilon)}{E(\varepsilon)}\tanh\frac{E(\varepsilon)}{2T} \, ,
\eeqy
where $g(\varepsilon)$ is the density of s.p. states (per unit energy) given by 
\beqy
\label{23}
g(\varepsilon)=\frac{1}{4\pi^2}\frac{\sqrt{\varepsilon}}{B_q^{3/2}}\, .
\eeqy
The subscript $\Lambda$ is to indicate that the integral has to be regularized by introducing a cutoff. We 
include here all s.p. states whose energy lies below $\lambda_q+\varepsilon_\Lambda$ where $\varepsilon_\Lambda$ is a 
pairing energy cutoff. 

In the weak-coupling approximation, i.e. $\Delta_q\ll \lambda_q$ and $\Delta_q\ll\varepsilon_\Lambda$,
it was shown in Ref.~\cite{cha10b} that the pairing gap at $T=0$ is approximately given by
\beqy
\label{24}
\Delta_q(0)=2\lambda_q\exp\left(\frac{1}{g(\lambda_q) v^{\pi\,q}} \right)\exp\biggl[\frac{1}{2}\Lambda\left(\frac{\varepsilon_\Lambda}{\lambda_q}\right)\biggr]
\eeqy
where 
\beqy
\label{25}
\Lambda(x)=\ln (16 x) + 2\sqrt{1+x}-2\ln\left(1+\sqrt{1+x}\right)-4\quad .
\eeqy
The Lagrange multiplier $\lambda_q$ is approximately given by the Fermi energy
\beqy
\label{26}
\lambda_q=B_q k_{{\rm F}q}^2 \, ,
\eeqy
with $k_{{\rm F}q}=(3\pi^2\rho_q)^{1/3}$. 
Note that these expressions were obtained by going beyond the usual ``weak-coupling approximation''
in which the density of s.p. states is taken as a constant. Even though this provides a good approximation 
in the case of conventional BCS superconductivity~\cite{bcs57}, it is less accurate in the nuclear context 
because many more states are involved in the pairing mechanism. 
The temperature-dependence of the pairing gap can be very well represented by~\cite{gor96}
\beqy
\Delta_q(T\leq T_{\rm c})\simeq \Delta_q(0) \sqrt{1-\left(\frac{T}{T_{\rm c}}\right)^{\delta}}\, ,
\eeqy
with $\delta\simeq3.23$ and the critical temperature is given by 
\beqy
T_c= \Delta_q(0)\frac{\exp(\zeta)}{\pi}\simeq 0.57 \Delta_q(0)\, .
\eeqy

Phenomenological pairing functionals whose parameters have been fitted to nuclei generally yield unrealistic 
pairing gaps in homogeneous nuclear matter~\cite{tak94,cha08}. 
Given the uncertainties regarding pairing correlations 
in nuclei, Garrido et al.~\cite{gar99} proposed to determine the parameters of the pairing strength in Eq.~(\ref{15}) by fitting the 
$^1S_0$ pairing gaps in infinite symmetric nuclear matter (SNM) as obtained by the realistic Paris potential in the BCS 
approximation. The pairing interaction between two nucleons inside a nucleus is thus assumed to be 
locally the same as the pairing interaction between two nucleons in infinite uniform matter. Even though the coupling 
to surface vibrations is expected to contribute to pairing~\cite{bri05} (see 
also Kamerdzhiev and Avdeenkov in this volume), a local 
pairing theory seems a reasonable first step (see Bulgac in this volume). 
The main difficulty of this approach is that because of the highly non-linear 
character of pairing correlations~\cite{dg04,zh10} it is very difficult to 
guess an appropriate functional form for $v^{\pi\,q}[\rho_n,\rho_p]$ 
(see also Lombardo in this volume). 
For instance, Margueron et al.~\cite{mar08} have recently shown that the parametric form~(\ref{15}) has to be generalized 
in order to reproduce the $^1S_0$ pairing gaps in both SNM and pure neutron matter (NeuM) as obtained from 
Brueckner calculations~\cite{cao06}. 

\begin{figure*}
\begin{center}
\includegraphics[scale=0.3,angle=-90]{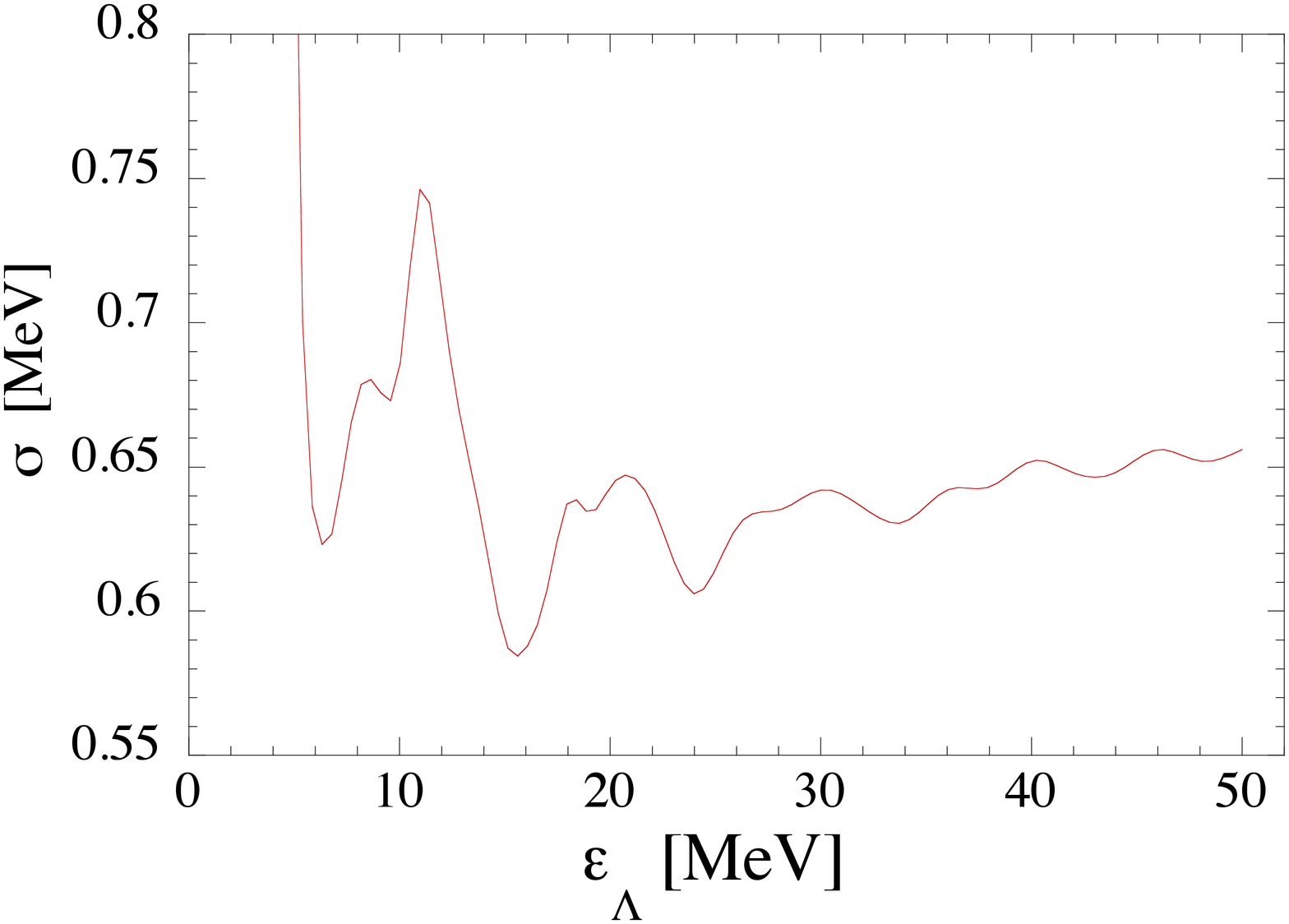}
\end{center}
\vskip -0.5cm
\caption{Root mean square deviation $\sigma$ between theoretical and experimental masses for some 260 known spherical or quasi-spherical 
nuclei for different values of the pairing cut-off energies. See text for more details.
}
\label{fig_rms}
\end{figure*}

Alternatively, Eq.~(\ref{24}) can be inverted to obtain the analytic expression of the pairing strength in terms of a given 
pairing gap function $\Delta_q(\rho_n,\rho_p)$
\beqy
\label{27}
v^{\pi\,q}[\rho_n,\rho_p]=-\frac{8\pi^2 B_q^{3/2} }{I_q(\rho_n,\rho_p)}
\eeqy
with 
\beqy
\label{28}
I_q=\sqrt{\lambda_q}\biggl[ 2\ln\left(\frac{2\lambda_q}{\Delta_q}\right)+ \Lambda\left(\frac{\varepsilon_\Lambda}{\lambda_q}\right) \biggr]\, .
\eeqy
The value of the pairing cutoff $\varepsilon_\Lambda$ is not completely arbitrary and could be fixed as follows. 
It has been argued~\cite{esb97,gar99} that in the limit $\rho\rightarrow0$, the pairing strength should coincide with the bare 
force in the $^1$S$_0$ channel, which in turn is determined by the experimental $^1$S$_0$ nucleon-nucleon phase shifts. However at 
very low densities, the pairing strength is simply given by 
\beqy
\label{29}
v^{\pi\,q}[\rho\rightarrow0]=-\frac{4\pi^2}{\sqrt{\varepsilon_\Lambda}}\left(\frac{\hbar^2}{2 M_q}\right)^{3/2} \, .
\eeqy 
The optimal value of the cutoff is thus found to be $\varepsilon_\Lambda\sim 7-8$ MeV (note that $\varepsilon_\Lambda$ is half 
the cutoff used in Ref.~\cite{esb97}). On the other hand, such a low cutoff may not be optimal in applications to finite nuclei. 
As seen in Fig.~\ref{fig_rms}, the root mean square (rms) deviation obtained with respect to about 260 known masses of (quasi-)spherical 
nuclei is found to oscillate as a function of the cutoff energy with clear minima lying around 7, 16 and 24 MeV. In this example, the 
initial EDF has the same characteristics as BSk21 with a pairing functional constrained on nuclear 
matter properties~\cite{gcp10}, as described in the next section. Note that for each value of the cut-off, the Skyrme 
interaction parameters were re-adjusted to minimize the rms deviation. In particular, we found systematically that global fits to 
nuclear masses favor $\varepsilon_\Lambda\sim 16$ MeV, a value which we adopted. Similar results were obtained when considering a 
traditional $\delta$-pairing force with or without the Bulgac-Yu regularization~\cite{go06}.

\section{Pairing in nuclei}

\begin{figure*}
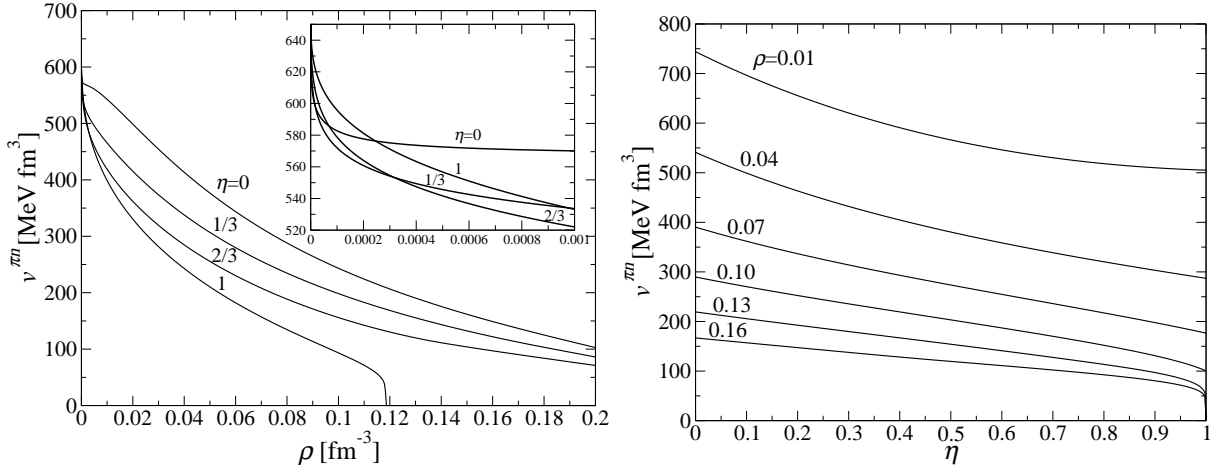

\begin{center}
\includegraphics[scale=0.3]{fig_vpi1}
\includegraphics[scale=0.3]{fig_vpi2}
\end{center}
\vskip -0.5cm
\caption{Neutron pairing strength as a function of the 
nucleon density $\rho=\rho_n+\rho_p$ and isospin asymmetry 
$\eta=(\rho_n-\rho_p)/\rho$.
}
\label{fig_vpi}
\end{figure*}

We have recently constructed a family of three nuclear EDFs, BSk19, BSk20 and 
BSk21~\cite{gcp10} based on Skyrme forces that are generalized in the sense
that they contain density-dependent generalizations of the usual $t_1$ and 
$t_2$ terms, respectively~\cite{cgp09}. The neutron-pairing functional was 
obtained from  Eqs.~(\ref{27})-(\ref{28}) using the $^1S_0$ pairing gaps both 
in SNM matter and NeuM, as obtained from Brueckner calculations including 
medium polarization effects~\cite{cao06}. The resulting pairing strength is shown 
in Fig.~\ref{fig_vpi}. The proton-pairing functional had the
same form but we allowed its strength to be different from the neutron-pairing 
strength in order to take account of Coulomb effects not included in the above 
nuclear matter calculations. Because of our neglect of polarization effects in 
odd nuclei due to our use of the equal-filling approximation~\cite{pmr08}, we also allowed 
each of these strengths to depend on whether there is an even or odd number of 
nucleons of the charge type in question\footnote{Note that the odd nucleon will
nevertheless contribute to the time-even fields.}. 
These extra degrees of freedom were taken into account by multiplying the 
neutron-pairing functional $v^{\pi\,q}[\rho_n, \rho_p]$, as determined 
by the nuclear-matter calculations that we have just described,
with renormalizing factors $f^{\pm}_q$, where $f^+_p, f^-_p$ and $f^-_n$ are 
free, density-independent parameters to be included in the mass fit, with 
$f^+_n$ set equal to $1$. 

The remaining parameters of the functionals were determined primarily by 
fitting essentially all measured nuclear masses. For this it was necessary to 
add two phenomenological corrections to the HFB ground-state energy: (i) a 
Wigner energy (which contributes significantly only for light nuclei or nuclei 
with $N$ close to $Z$) and (ii) a correction for the spurious rotational and 
vibrational collective energy. 
However, in fitting the mass data we simultaneously constrained the functional to 
fit the zero-temperature equation of state (EOS)
of NeuM, as determined by three different many-body calculations 
using realistic two- and three-nucleon forces. 
Finally, we imposed on these EDFs the supplementary constraints of i) eliminating all 
spurious spin-isospin instabilities in nuclear matter both at zero and finite temperatures, 
at all densities found in neutron stars and supernova cores~\cite{cgp09,cg10,gcp10}, ii) obtaining a 
qualitatively realistic distribution of the potential energy among the four spin-isospin channels 
in nuclear matter, iii) ensuring that the isovector effective mass is smaller than the isoscalar 
effective mass, as indicated by both experiments and many-body calculations and iv) restricting
the incompressibility coefficient of SNM to lie in the range $K_v=240\pm 10$ MeV, as inferred 
from breathing-mode measurements. 

The introduction of the unconventional Skyrme terms allowed us to satisfy all 
these constraints and at the same time fit the 2149 measured masses of nuclei 
with $N$ and $Z \ge$ 8 given in the 2003 Atomic Mass Evaluation 
(AME)~\cite{audi03} with an rms deviation as low as 0.58 MeV for all three EDFs. 
Incidentally, our EDFs are found to be consistent with measurements of the high-density 
pressure of SNM deduced from heavy-ion collisions~\cite{dan02}, even though they were 
not directly fitted to the EOS of SNM.

\section{Pairing in neutron-star crusts}

Because of the precision fit to masses and the constraints on both the EOS and the $^1S_0$ 
pairing gaps in NeuM, our recently developed EDFs BSk19, BSk20 and BSk21~\cite{gcp10} are particularly 
well-suited for describing the inner crust of neutron stars.  

The HFB equations~(\ref{8}) have been already solved in neutron-star 
crusts using the so-called Wigner-Seitz (W-S) approximation according to which
the crust is divided into a set of independent spheres centered 
around each lattice site~\cite{bal07,gri10,pas10} (see also Barranco et al. in this 
volume). However this way of implementing 
the HFB method can only be reliably applied in the shallowest region of the inner 
crust where nuclear clusters are very far apart~\cite{cha07}. In order to investigate pairing correlations 
in the densest part of the crust, we have applied the band theory of solids, which takes into 
account both short- and long-range correlations~\cite{cha10}. The band theory relies on the assumption that the solid crust 
can be treated as a perfect crystal, which is a reasonable approximation for cold non-accreting neutron 
stars~\cite{lrr}. According to the Floquet-Bloch theorem, the q.p. wave function must obey 
the following boundary conditions~\cite{mat76} for any lattice translation vectors $\pmb{\ell}$
\beqy\label{30}
\psi^{(q)}_{1\alpha\pmb{k}}(\pmb{r}+\pmb{\ell}, \sigma)=\exp({\rm i} \pmb{k}\cdot\pmb{\ell})\,\psi^{(q)}_{1\alpha\pmb{k}}(\pmb{r}, \sigma)\nonumber\\
 \psi^{(q)}_{2\alpha\pmb{k}}(\pmb{r}+\pmb{\ell}, \sigma)=\exp({\rm i} \pmb{k}\cdot\pmb{\ell})\,\psi^{(q)}_{2\alpha\pmb{k}}(\pmb{r}, \sigma) 
\eeqy
where $\alpha$ is the band index (principal quantum number) and $\pmb{k}$ the Bloch wave vector. 
This formalism naturally incorporates the local 
rotational symmetries around the nuclear clusters as well as the translational symmetry of the lattice, 
thus describing consistently both clusters and superfluid neutrons. Note that this formalism also includes 
infinite homogeneous matter as the limiting case of an ``empty'' lattice. The band theory therefore allows for a unified 
treatment of all regions of a neutron star. 

In the deep layers of the inner crust of a neutron star, where spatial inhomogeneities are small, further 
simplifications can be made. In the decoupling approximation, the q.p. wavefunction is expressed 
in terms of the s.p. wavefunctions $\varphi^{(q)}_{\alpha\pmb{k}}$ as  
\beqy
\label{31}
\psi^{(q)}_{1\alpha \pmb{k}}(\pmb{r},\sigma) = U^{(q)}_{\alpha \pmb{k}}\, \varphi^{(q)}_{\alpha \pmb{k}}(\pmb{r},\sigma)  \, , \hskip0.5cm \psi^{(q)}_{2\alpha \pmb{k}}(\pmb{r},\sigma)=V_{\alpha \pmb{k}}^{(q)}\, \varphi^{(q)}_{\alpha \pmb{k}}(\pmb{r},\sigma) \, .
\eeqy
The HFB equations can then be readily solved, and the q.p. energies are given by
\beqy
\label{32}
E^{(q)}_{\alpha\pmb{k}}=\sqrt{(\varepsilon^{(q)}_{\alpha\pmb{k}}-\lambda_q)^2 +(\Delta^{(q)}_{\alpha\pmb{k}})^2}
\eeqy
where $\varepsilon^{(q)}_{\alpha\pmb{k}}$ are the s.p. energies and $\Delta^{(q)}_{\alpha\pmb{k}}$ are solutions of the 
anisotropic multi-band BCS gap equations~\cite{cha10}
\beqy
\label{33}
\Delta^{(q)}_{\alpha\pmb{k}} = - \frac{V}{2} \sum_{\beta} \int \frac{{\rm d}^3 \pmb{k^\prime}}{(2\pi)^3} V^{(q)}_{\alpha \pmb{k}\beta \pmb{k^\prime}}
 \frac{\Delta^{(q)}_{\beta\pmb{k^\prime}}}{E^{(q)}_{\beta \pmb{k^\prime}} }\tanh \frac{E^{(q)}_{\beta \pmb{k^\prime}}}{2T} \, .
\eeqy
with 
\beqy
\label{34}
V^{(q)}_{\alpha \pmb{k}\beta \pmb{k^\prime}} =\int_{\rm WS} {\rm d}^3r\, v^{\pi\, q}[\rho_n(\pmb{r}),\rho_p(\pmb{r})]\, 
|\varphi^{(q)}_{\alpha\pmb{k}}(\pmb{r})|^2 |\varphi^{(q)}_{\beta\pmb{k^\prime}}(\pmb{r})|^2\, .
\eeqy
The subscript WS indicates that the integral has to be taken inside the W-S cell. Finally the amplitudes of the 
q.p. wavefunction are given by 
\beqy
\label{35}
U^{(q)}_{\alpha \pmb{k}}= \frac{1}{\sqrt{2}}\left(1+\frac{\varepsilon^{(q)}_{\alpha\pmb{k}}-\lambda_q}{E^{(q)}_{\alpha\pmb{k}}}\right)^{1/2} \, ,
\eeqy
\beqy
\label{36}
V^{(q)}_{\alpha \pmb{k}}= \frac{1}{\sqrt{2}}\left(1-\frac{\varepsilon^{(q)}_{\alpha\pmb{k}}-\lambda_q}{E^{(q)}_{\alpha\pmb{k}}}\right)^{1/2} \, .
\eeqy

\begin{figure*}
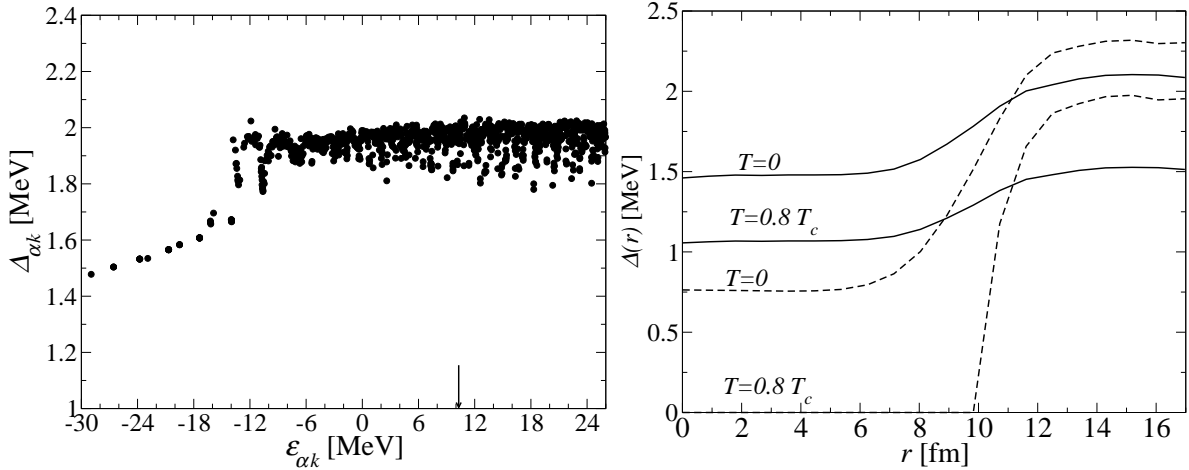

\begin{center}
\includegraphics[scale=0.3]{fig_gaps}
\includegraphics[scale=0.3]{fig_pairfield}
\end{center}
\vskip -0.5cm
\caption{Neutron superfluidity in neutron-star crust at average baryon 
density $\bar \rho = 0.06$ fm$^{-3}$ with BSk21. 
The critical temperature is found to be equal to $T_{\rm c}=1.1$ MeV.
Left panel: pairing gaps vs single-particle energies at $T=0$. 
The arrow indicates the position of the Fermi energy.
Right panel: pairing field in the W-S cell obtained from the 
HFB equations (solid line) and from the LDA (dashed line) for two different temperatures.
}
\label{fig_pair}
\end{figure*}

We have solved Eq.~(\ref{33}) for neutrons in the deep region of neutron-star crusts as described in Ref.~\cite{cha10} using our latest
BSk21 EDF (which is strongly favored by the most recent atomic mass data while being also consistent with what is now known 
about neutron-star masses~\cite{cfpg11}). Due to the presence of the nuclear clusters, 
neutrons belonging to different bands and having different Bloch wave vectors feel different pairing interactions thus leading 
to a dispersion of the neutron pairing gaps $\Delta^{(n)}_{\alpha\pmb{k}}$ of a few hundred keV around the Fermi level, 
as shown in Fig.~\ref{fig_pair}. 
The critical temperature is found to be very weakly dependent on the cutoff, as can be seen in Table~\ref{tab1}
(note that $\varepsilon_\Lambda$ was varied while using the \emph{same} Skyrme functional BSk21). 

Because the neutron superfluid coherence length is much larger than the size of the clusters, proximity effects are
very strong. As a result, pairing correlations are substantially enhanced inside clusters while they are reduced in the 
intersticial region, leading to a smooth spatial variation of the pairing field. The local density approximation (LDA), 
whereby the neutron pairing field $\Delta_n(\pmb{r})$ is assumed to be locally the same as that in uniform nuclear 
matter for the neutron density $\rho_n(\pmb{r})$ and proton density $\rho_p(\pmb{r})$, strongly overestimates the 
spatial variations of the pairing field. The discrepancies are particularly large inside clusters where the LDA 
incorrectly predicts a quenching of pairing correlations, especially for temperatures close to the critical temperature
as illustrated in Fig.~\ref{fig_pair}. 
This analysis shows that a consistent treatment of both unbound neutrons and nucleons bound in clusters is essential 
for a realistic description of pairing correlations in 
neutron-star crusts. 

\begin{table}
\centering
\caption{Cut-off dependence of the critical temperature of neutron superfluidity in neutron star crusts at density 
$\bar \rho=0.06$ fm$^{-3}$ with BSk21. See text for further details.
}
\label{tab1}
\vspace{.5cm}
\begin{tabular}{|c|c|c|c|c|c|}
\hline
$\varepsilon_\Lambda$ [MeV] & 2 & 4 & 8 & 16  & 32  \\ 
\hline
$T_{\rm c}$ [MeV] & 1.10 & 1.06 & 1.07 & 1.11 & 1.16 \\ 
\hline
\end{tabular}
\end{table}

Despite the absence of viscous drag at $T=0$, the solid crust can still resist the flow of the neutron superfluid 
due to non-dissipative entrainment effects. These effects have been systematically studied in all regions of the inner 
crust of a cold non-accreting neutron star~\cite{cha12}. In particular, it has been found that in some layers of the 
inner crust, almost all neutrons are entrained by clusters. These results suggest that a revision of the interpretation 
of many observable astrophysical phenomena like pulsar glitches might be necessary~\cite{cc06}.

\section{Conclusions}

The nuclear EDF theory opens the way to a unified description of the nuclear pairing phenomenon in various 
systems, from atomic nuclei to neutron stars. 

The Brussels-Montreal EDFs based on generalized Skyrme EDFs supplemented by a microscopic local pairing EDF 
yield an excellent fit to essentially all experimental nuclear mass data with rms deviations falling below 
0.6 MeV, while reproducing at the same time many-body calculations in infinite homogeneous nuclear matter using realistic forces. 
For this reason, these EDFs are particularly well-suited for studying pairing correlations in the inner crust of 
neutron stars, where nuclear clusters are expected to coexist with a neutron superfluid. 

Despite these successes, a number of open issues like for instance neutron-proton pairing or the 
contribution of surface vibrations to pairing call for more elaborate pairing EDFs.

\end{document}